# Digraphs with Distinguishable Dynamics under the Multi-Agent Agreement Protocol

M. Amin Rahimian, Amir Ajorlou, and Amir G. Aghdam

**ABSTRACT**

In this work the ability to distinguish digraphs from the output response of some observing agents in a multi-agent network under the agreement protocol has been studied. Given a fixed observation point, it is desired to find sufficient graphical conditions under which the failure of a set of edges in the network information flow digraph is distinguishable from another set. When the latter is empty, this corresponds to the detectability of the former link set given the response of the observing agent. In developing the results, a powerful extension of the all-minors matrix tree theorem in algebraic graph theory is proved which relates the minors of the transformed Laplacian of a directed graph to the number and length of the shortest paths between its vertices. The results reveal an intricate relationship between the ability to distinguish the responses of a healthy and a faulty multi-agent network and the inter-nodal paths in their information flow digraphs. The results have direct implications for the operation and design of multi-agent systems subject to multiple link losses. Simulations and examples are presented to illustrate the analytic findings.

The authors are with the Department of Electrical and Computer Engineering, Concordia University, 1455 De Maisonneuve Blvd. W., Montréal, Québec, H3G 1M8, Canada (email: aghdam@ece.concordia.ca). This work has been supported by the Natural Sciences and Engineering Research Council of Canada (NSERC) under grant RGPIN-262127-12.

# I. INTRODUCTION

Multi-agent network systems consist of a group of dynamic agents, which interact according to a given information flow structure [10]. These systems have found promising applications in areas such as formation control of satellite clusters and motion coordination of robots [6, 12]. Distributed and cooperative control for these networked dynamic systems employs various concepts from different fields including parallel processing, distributed algorithms, control, and estimation [2]. Popular research problems related to multi-agent network control include connectivity, containment, consensus, rendezvous, formation, flocking, and controllability [13, 5, 19, 27, 9]. Such cooperative dynamics over a network may be strongly affected by the network failures, and this has motivated the study of network dynamics following the removal of some links or nodes [7, 16], as well as the related studies focusing on switching network topologies [9]. By and large, the study of failures is an important topic in network science and it has various practical implications [8].

Agreement protocols have been extensively investigated in the recent literature as a fundamental evolution law for multi-agent networks in both continuous and discrete-time, using probabilistic and deterministic models [14, 15, 20, 21, 22, 25]. Earlier results on adjacency-based agreement rules can be traced back to Vicsek's model [24]. In [26], the authors calculate the single-input single-output transfer functions between pairs of nodes acting as the control input and measurement nodes for cyclic consensus systems. They also point out the interesting implications of the length of the shortest path between the input and measurement nodes on the stability margins and the relative degree of the calculated transfer functions. The issues of identification and infiltration in consensus-type networks is investigated in [4], where the authors consider additional observation and excitation nodes and make use of the available system identification techniques to derive bounds on the eigenvalues of the network graph.

This paper focuses on the cooperative control of a multi-agent system under the linear agreement protocol and subject to multiple communication link failures. The mathematical characterization of simultaneous link failures in the paper leads to useful design guidelines for realization of reliable and fault-tolerant multi-agent networks. Detection and isolation of faults are crucial to the cooperative and reliable control of multi-agent networks, and the chief aim of this paper is to address these two important concepts. The results are therefore of both theoretical and practical interest.

The remainder of this paper is organized as follows. Section II gives some preliminaries on sets and graph theory, and introduces the notation that is used throughout the paper. The background provided in this section is then used in Section III, where topological conditions for producing distinguishable dynamics in the observed response of an agent are investigated. Next, in Section IV, the detectability of link failures is considered as a special case and sufficient conditions are derived accordingly. Finally, concluding remarks are provided in Section V.

# II. PRELIMINARIES AND NOTATION

Throughout the paper, $\varnothing$ is the empty set, $\mathbb{N}$ denotes the set of all natural numbers, $\mathbb{R}$ denotes the set of all real numbers, and $\mathbb{C}$ denotes the set of all complex numbers. Also, the set of integers $\{1, 2, \ldots, k\}$ is denoted by $\mathbb{N}_k$, and any other set is represented by a curved capital letter. The cardinality of a set $\mathcal{X}$, which is the number of its elements, is denoted by $|\mathcal{X}|$. The difference of two sets $\mathcal{X}$ and $\mathcal{Y}$ is denoted by $\mathcal{X} \backslash \mathcal{Y}$ and is defined as $\{x; x \in \mathcal{X} \wedge x \notin \mathcal{Y}\}$, where $\wedge$ is the logical conjunction. Matrices are represented by capital letters, vectors are expressed by boldface lower-case letters, and the superscript $^T$ denotes the matrix transpose. Moreover, $I$, $\mathbf{1}$, and $\mathbf{0}$ denote the identity matrix and column vectors with all one and all zero entries, respectively; and their dimensions are clear from the context. The determinant of a matrix $M$ is denoted by $\det(M)$, while $[M]_{ij}$ indicates the element of $M$ which is located at its $i-$th row and $j-$th column.

## 2.1. Directed Graphs and the Associated Algebraic Entities

A directed graph or *digraph* is defined as an ordered pair of sets: $\mathcal{G} = (\mathcal{V}, \mathcal{E})$, where $\mathcal{V} = \{\nu_1, \ldots, \nu_n\}$ is a set of $n = |\mathcal{V}|$ vertices and $\mathcal{E} \subseteq \mathcal{V} \times \mathcal{V}$ is a set of directed edges. In the graphical representations, each edge $\epsilon := (\tau, \nu) \in \mathcal{E}$ is denoted by a directed arc from vertex $\tau \in \mathcal{V}$ to vertex $\nu \in \mathcal{V}$. Vertices $\nu$ and $\tau$ are referred to as the *head* and *tail* of the edge $\epsilon$, respectively. Given a set of vertices $\mathcal{X} \subset \mathcal{V}$, the set of all edges for which the tails belong to $\mathcal{X}$ but the heads do not, is termed the out-cut of $\mathcal{X}$, and is denoted by $\partial_{\mathcal{G}}^+ \mathcal{X} \subset \mathcal{E}$. Similarly, the set of all edges for which the heads belong to $\mathcal{X}$ but the tails do not, is referred to as the in-cut of $\mathcal{X}$, and is denoted by $\partial_{\mathcal{G}}^{-}\mathcal{X} \subset \mathcal{E}$. The cardinality of $\partial_{\mathcal{G}}^{-}\mathcal{X}$ is called the in-degree of $\mathcal{X}$, and is characterized as $d_{\mathcal{G}}^{-}\mathcal{X} = |\partial_{\mathcal{G}}^{-}\mathcal{X}|$. Notice that the definition of $\mathcal{E}$ does not allow for the existence of parallel arcs in the graphical representation of $\mathcal{G}$. In other words, if two edges share the same pair of head and tail, then they are identical.

Given an integer $k \in \mathbb{N}_{n-2}$, a permutation $(\alpha_1, \alpha_2, \ldots, \alpha_k)$ of all members of the set $\mathbb{N}_k$, and two vertices $\tau, \nu \in \mathcal{V}$, a sequence of distinct edges of the form $\mathcal{P} := (\tau, \nu_{\alpha_1}), (\nu_{\alpha_1}, \nu_{\alpha_2}), \ldots, (\nu_{\alpha_{k-1}}, \nu_{\alpha_k}), (\nu_{\alpha_k}, \nu)$ is called a $\tau\nu$ *path* with *length* $k + 1$ if for any two edges $(\bar{\tau}, \bar{\nu}), (\hat{\tau}, \hat{\nu})$ of this sequence, $\bar{\nu} \neq \hat{\nu} \longleftrightarrow \bar{\tau} \neq \hat{\tau}$. For a given vertex $\nu \in \mathcal{V}$, a $\nu\nu$ path is called a *directed circle*. Moreover, the shortest length for all $\tau\nu$ paths is referred to as the *distance from $\tau$ to $\nu$*, and is denoted by $\mathrm{d}(\tau, \nu)$. Likewise, the number of $\tau\nu$ paths with length $k$, denoted by $\mathrm{c}_k(\tau, \nu)$, is called the $k$-*th order topological connectivity* of $\tau$ to $\nu$. Notably, $\mathrm{c}_k(\tau, \nu) = 0$ for $k < \mathrm{d}(\tau, \nu)$, and by convention $\mathrm{d}(\nu, \nu) = 0$, $\mathrm{c}_0(\nu, \nu) = 1$, and if $\forall k \in \mathbb{N}, \mathrm{c}_k(\tau, \nu) = 0$, then $\mathrm{d}(\tau, \nu) = \infty$.

For a given digraph $\mathcal{G} = (\mathcal{V}, \mathcal{E})$ and two vertices $\nu_i, \nu_j \in \mathcal{V}$, the *edge-index* of $(\nu_i, \nu_j)$ is defined as a $|\mathcal{V}| \times |\mathcal{V}|$ matrix whose only non-zero element is 1, which is located at its $j$-th row and $i$-th column. This matrix is represented by $\Gamma((\nu_i, \nu_j))$. Similarly, the *vertex-index* of any $\nu_i \in \mathcal{V}$ is defined as a $|\mathcal{V}| \times 1$ column vector whose only non-zero element is 1, which is located at its $i$-th row. This vector is denoted by $\boldsymbol{\sigma}(\nu_i)$. The adjacency matrix of $\mathcal{G}$ is given by $A(\mathcal{G}) = \sum_{\epsilon \in \mathcal{E}} \Gamma(\epsilon)$, its *degree matrix* is defined as $\Delta(\mathcal{G}) = \sum_{\nu \in \mathcal{V}} d_{\mathcal{G}}^{-}\{\nu\}\Gamma((\nu, \nu))$, and the corresponding in-degree graph Laplacian is given by $\mathcal{L}(\mathcal{G}) = \Delta(\mathcal{G}) - A(\mathcal{G})$.

For a given digraph $\mathcal{G} = (\mathcal{V}, \mathcal{E})$, a vertex $\tau \in \mathcal{V}$ is called an *out-branching root* if there exists a $\tau\nu$ path for every $\nu \in \mathcal{V} \setminus \{\tau\}$. Furthermore, a digraph $\tilde{\mathcal{G}} = (\mathcal{V}, \tilde{\mathcal{E}}), \tilde{\mathcal{E}} \subseteq \mathcal{E}$ with an out-branching root $\tilde{\tau} \in \mathcal{V}$ is a $\tilde{\tau}$-*rooted spanning out-branching* of $\mathcal{G}$ if $\tilde{\mathcal{G}}$ does not contain any directed circle and $\bar{\tau} \neq \hat{\tau} \longrightarrow \bar{\nu} \neq \hat{\nu}$ for any $\{(\bar{\tau}, \bar{\nu}), (\hat{\tau}, \hat{\nu})\} \subset \tilde{\mathcal{E}}$.

The following Theorem is known as the *all-minors matrix tree theorem* in algebraic graph theory, and is a generalization of the Matrix Tree Theorem for the case of directed graphs [23].

**Theorem 1** *For a given digraph $\mathcal{G} = (\mathcal{V}, \mathcal{E})$ and a vertex $\nu_i \in \mathcal{V}$, the number of $\nu_i$-rooted spanning out-branchings of $\mathcal{G}$ is equal to any cofactor in the $i$-th row of $\mathcal{L}(\mathcal{G})$.*

The next lemma is a significant extension to Theorem 1. This lemma relates the minors of the Laplace-transformed graph Laplacian to the length and number of inter-nodal paths in the digraph. Its proof uses the permutation expansion of the determinants and follows the steps of the combinatorial proof of the all-minors matrix tree theorem in [3], by relating the sum of the products of the entries of the graph Laplacian over some class of permutations to the topological connectivity of the graph nodes.

**Lemma 1** *Given a digraph $\mathcal{G} = (\mathcal{V}, \mathcal{E})$, let $H = sI + \mathcal{L}(\mathcal{G})$, $s \in \mathbb{C}$. For any $i, j \in \mathbb{N}_{|\mathcal{V}|}$ and $i \neq j$, define $C_{ij}$ as the matrix that results from removing the $i$-th row and $j$-th column of $H$. Then $\det(C_{ij})$ is a polynomial of degree $|\mathcal{V}| - \mathrm{d}(\nu_i, \nu_j) - 1$ in $s$, whose leading coefficient has an absolute value equal to $\mathrm{c}_{\mathrm{d}(\nu_i, \nu_j)}(\nu_i, \nu_j)$.*

**Proof:** $\det(C_{ij})$ can be written as a summation over the product of all possible permutations on the choice of $|\mathcal{V}| - 1$ matrix elements [11], as given below:

$$\det(C_{ij}) = \sum_{\psi \in \Xi_{|\mathcal{V}|-1}} (-1)^{\iota(\psi)} \prod_{k=1}^{|\mathcal{V}|-1} [C_{ij}]_{k\psi(k)}, \tag{1}$$

where $\Xi_{|\mathcal{V}|-1}$ is the finite group formed by the $(|\mathcal{V}| - 1)!$ permutations on the set $\mathbb{N}_{|\mathcal{V}|-1}$. Moreover, for a permutation $\psi(.)$ on the set $\mathbb{N}_{|\mathcal{V}|-1}$, $\iota(\psi)$ denotes its length and is defined as the number of pairs $i, j \in \mathbb{N}_{|\mathcal{V}|-1}$, $i < j$, such that $\psi(i) > \psi(j)$. Consider the case where $\mathrm{d}(\nu_i, \nu_j) = 1$, which implies that there exists an edge $(\nu_i, \nu_j) \in \mathcal{E}$. Hence, $[\mathcal{L}(\mathcal{G})]_{ji} = -1$ and $\det(C_{ij})$ is a polynomial of degree $|\mathcal{V}| - 2$ in $s$. Moreover, the only term in $\det(C_{ij})$ which includes $s^{|\mathcal{V}|-2}$ in the right-hand side of (1) is the one corresponding to the elements $[H]_{rr}, r \in \mathbb{N}_{|\mathcal{V}|} \setminus \{i, j\}$ and $[H]_{ji} = [\mathcal{L}(\mathcal{G})]_{ji}$. Accordingly, (1) can be rewritten as:

$$\det(C_{ij}) = (-1)^{\kappa}[\mathcal{L}(\mathcal{G})]_{ji} \left( \prod_{\nu \in \mathcal{V} \setminus \{\nu_i, \nu_j\}} (s + d_{in}(\nu)) \right) + w_1(s) = -(-1)^{\kappa} s^{|\mathcal{V}|-2} + w_2(s), \tag{2}$$

for some $\{w_1(s), w_2(s)\} \subset \mathcal{W}_{|\mathcal{V}|-3}(s)$, where $\mathcal{W}_{|\mathcal{V}|-3}(s)$ denotes the set of all polynomials in $s$ with degrees less than or equal to $|\mathcal{V}| - 3$ and a permutation length $\kappa \in \mathbb{N}$ that depends on $i$, $j$ and $|\mathcal{V}|$. This corroborates (1) for $d(\nu_i, \nu_j) = 1$. Next, let $[\mathcal{L}(\mathcal{G})]_{mn} = L_{mn}, \{m, n\} \subset \mathbb{N}_{|\mathcal{V}|}$ and consider the expression for $\det(C_{ij})$ in the general case of $d(\nu_i, \nu_j) \in \mathbb{N}_{|\mathcal{V}|-1} \setminus \{1\}$. From $d(\nu_i, \nu_j) > 1$, it follows that $[\mathcal{L}(\mathcal{G})]_{ji} = 0$ and (1) can be expanded as:

$$\det(C_{ij}) = (-1)^{\kappa_2} \sum_{r=1, r \neq i,j}^{|\mathcal{V}|} \left( L_{jr} L_{ri} \prod_{k=1, k \neq i,j,r}^{|\mathcal{V}|} (s + d_{in}(\nu_k)) \right) +$$

$$(-1)^{\kappa_3} \sum_{r=1, r \neq i,j}^{|\mathcal{V}|} \sum_{t=1, t \neq i,j,r}^{|\mathcal{V}|} \left( L_{jr} L_{rt} L_{ti} \prod_{k=1, k \neq i,j,r,t}^{|\mathcal{V}|} (s + d_{in}(\nu_k)) \right) + \ldots, \quad (3)$$

for some $\kappa_l$, $l \in \mathbb{N}_{|\mathcal{V}|-2} \setminus \{1\}$, which depend on the indices $i$ and $j$ and vary with the length of the permutations. To express (2) in a more compact manner, let $\Theta = \{\theta_1, \theta_2, \ldots, \theta_{|\Theta|}\}$, $\Pi = \{\Theta; \Theta \subseteq \mathbb{N}_{|\mathcal{V}|} \setminus \{i, j\} \wedge \Theta \neq \varnothing\}$ so that $|\Pi| = 2^{|\mathcal{V}|-2} - 1$, and let $\Xi_{|\Theta|}$ be the finite group formed by the $(|\Theta|)!$ permutations on the set $\Theta$. Furthermore, for $\Theta \in \Pi$ and $\psi \in \Xi_{|\Theta|}$ define:

$$S(\Theta) = \prod_{k \in \mathbb{N}_{|\mathcal{V}|} \setminus (\{i,j\} \cup \Theta)} (s + d_{in}(\nu_k)), \quad (4a)$$

$$P(\Theta, \psi) = L_{j\theta_{\psi(1)}} \left( \prod_{l=1}^{|\Theta|-1} L_{\theta_{\psi(l)} \theta_{\psi(l+1)}} \right) L_{\theta_{\psi(|\Theta|)} i}. \quad (4b)$$

Using (4), for $d(\nu_i, \nu_j) > 1$, (3) can be rewritten as:

$$\det(C_{ij}) = \sum_{\Theta \in \Pi} (-1)^{\kappa_{|\Theta|}} \left( \sum_{\psi \in \Xi_{|\Theta|}} P(\Theta, \psi) \right) S(\Theta), \quad (5)$$

for some $\kappa_{|\Theta|} \in \mathbb{N}$ that depend on $i$ and $j$. Breaking the first summation over $|\Theta| = k$ for $k = 1, \ldots, |\mathcal{V}| - 2$ and using the notation $\Pi_k = \{\Theta; \Theta \subset \mathbb{N}_{|\mathcal{V}|} \setminus \{i, j\} \wedge |\Theta| = k\}$ in (5) yields:

$$\det(C_{ij}) = \sum_{k=1}^{|\mathcal{V}|-2} (-1)^{\kappa_k(i,j)} \sum_{\Theta \in \Pi_k} \left( \sum_{\psi \in \Xi_k} P(\Theta, \psi) \right) S(\Theta). \quad (6)$$

Next, note that given $k \in \mathbb{N}_{|\mathcal{V}|}$, two vertices $\{\nu_i, \nu_j\} \subset \mathcal{V}$ and $|\{\nu_i, \nu_j\}| = 2$, the $k$-th order topological connectivity of $\nu_j$ to $\nu_i$ can be expressed in terms of $[\mathcal{L}(\mathcal{G})]_{mn} = L_{mn}, \{m, n\} \subset \mathbb{N}_{|\mathcal{V}|}$ as follows:

$$c_k(\nu_j, \nu_i) = (-1)^k \times \sum_{\Theta \in \Pi_{k-1}} \left( \sum_{\psi \in \Xi_{k-1}} L_{i\theta_{\psi(1)}} \left( \prod_{l=1}^{k-2} L_{\theta_{\psi(l)} \theta_{\psi(l+1)}} \right) L_{\theta_{\psi(k-1)} j} \right), \quad (7)$$

where $\Xi_{k-1}$ is the finite group formed by the $(k-1)!$ permutations on the set $\mathbb{N}_{k-1}$, $\Theta = \{\theta_1, \theta_2, \ldots, \theta_{k-1}\} \subset \mathbb{N}_{|\mathcal{V}|} \setminus \{i, j\}$, and $\Pi_{k-1} = \{\Theta; \Theta \subset \mathbb{N}_{|\mathcal{V}|} \setminus \{i, j\} \wedge |\Theta| = k - 1\}$. Now from (7) and (4b) it follows that:

$$c_k(\nu_i, \nu_j) = (-1)^k \sum_{\Theta \in \Pi_{k-1}} \left( \sum_{\psi \in \Xi_{k-1}} P(\Theta, \psi) \right). \quad (8)$$

The fact that $S(\Theta)$ is a polynomial of degree $|\mathcal{V}| - |\Theta| - 2$ in $s$ with leading coefficient 1, together with (2), (6) and (8), leads to:

$$\det(C_{ij}) = K + \sum_{k=1}^{|\mathcal{V}|-2} (-1)^{\kappa_k(i,j)} c_k(\nu_i, \nu_j) w_{|\mathcal{V}|-k-1}(s), \quad (9)$$

where $w_l(s), l \in \mathbb{N}_{|\mathcal{V}|-2}$ are degree $l$ polynomials in $s$ whose leading coefficients are unity, and $|K| \in \mathbb{N}$ is a constant. The proof follows immediately from (9) and upon noting that $c_k(\nu_i, \nu_j) = 0$ for $k < d(\nu_i, \nu_j)$. ∎

## 2.2. Multi-Agent Systems under the Agreement Protocol

Consider a multi-agent system comprised of a set $\mathcal{S} = \{x_1, x_2, \ldots, x_n\}$ of $n$ single integrator agents, where $x_i$, $i \in \mathbb{N}_n$ is the state of agent $i$, which is assumed to be scalar. Further, for all $i \in \mathbb{N}_n$ assume that the control input of agent $i$ is constructed according to the following nearest neighbor law:

$$\dot{x}_i(t) = \sum_{(\nu_j, \nu_i) \in \partial_\mathcal{G}^- \{\nu_i\}} (x_j(t) - x_i(t)),\ t > 0. \tag{10}$$

The interaction structure between the agents in (10) can be described by a directed information flow graph $\mathcal{G} = (\mathcal{V}, \mathcal{E})$, where each vertex $\nu \in \mathcal{V}$ corresponds to an agent $x \in \mathcal{S}$ and $|\mathcal{V}| = n$. A directed edge from vertex $\nu_k$ to vertex $\nu_i$ implies that the term $x_k(t) - x_i(t)$ appears in the control law of the agent $x_i$ given by (10). As discussed in Section 3.2 of [10], the existence of an out-branching root $\tau \in \mathcal{V}$ is a necessary and sufficient condition for the states in (10) to converge to a common value. This correspondence is often referred to as the *agreement protocol*.

For a multi-agent system $\mathcal{S}$ and its associated digraph $\mathcal{G} = (\mathcal{V}, \mathcal{E})$, the in-degree graph Laplacian can be used to represent the dynamic equations in (10) in compact matrix form as follows:

$$\dot{\mathbf{x}}(t) = -\mathcal{L}(\mathcal{G})\mathbf{x}(t),\ t > 0, \tag{11}$$

where $\mathbf{x}(t) = (x_1(t), x_2(t), \ldots, x_n(t))^T$. The matrix exponential solution to (11) is then expressed as:

$$\mathbf{x}(t) = e^{-\mathcal{L}(\mathcal{G})t}\mathbf{x}(0),\ t \geqslant 0, \tag{12}$$

and for a particular agent $x_i \in \mathcal{S}$ represented by the vertex $\nu_i \in \mathcal{V}$, the temporal evolution of its state is given by:

$$x_i(t) = \boldsymbol{\sigma}(\nu_i)^T e^{-\mathcal{L}(\mathcal{G})t}\mathbf{x}(0),\ t \geqslant 0. \tag{13}$$

## III. DIGRAPHS WITH DISTINGUISHABLE DYNAMICS

The notion of distinguishable digraphs is formally defined next. This notion provides a mathematical characterization of link failures according to the state of an agent after the failure of a certain set of links in the information flow structure, and it is motivated by the fact that different link failures in the original digraph $\mathcal{G} = (\mathcal{V}, \mathcal{E}_1 \cup \mathcal{E}_2)$ can lead to distinct digraphs $\mathcal{G}_1 = (\mathcal{V}, \mathcal{E}_1)$ and $\mathcal{G}_2 = (\mathcal{V}, \mathcal{E}_2)$ which share the same set of vertices $\mathcal{V}$ but have different sets of edges $\mathcal{E}_1$ and $\mathcal{E}_2$.

**Definition 1** *Consider a multi-agent system $\mathcal{S}$ and two distinct digraphs $\mathcal{G}_1 = (\mathcal{V}, \mathcal{E}_1)$ and $\mathcal{G}_2 = (\mathcal{V}, \mathcal{E}_2)$ associated with it. Digraphs $\mathcal{G}_1$ and $\mathcal{G}_2$ are said to be* distinguishable *from an agent $x \in \mathcal{S}$ if there exists $\mathbf{x}(0) \in \mathbb{R}^{|\mathcal{V}|}$ such that $x^{\mathcal{G}_1}(t) - x^{\mathcal{G}_2}(t) \not\equiv 0$, $t > 0$, where $x^{\mathcal{G}_1}(t)$ and $x^{\mathcal{G}_2}(t)$ denote the state of the agent $x$ calculated in the digraphs $\mathcal{G}_1$ and $\mathcal{G}_2$, respectively, with the same initial condition $\mathbf{x}(0)$.*

**Example 1** *Consider the digraph $\mathcal{G}_1$ depicted in Fig. 1, and let $\mathcal{G}_2$ denote the digraph obtained by removing the edge $(\nu_3, \nu_2)$. The in-degree graph Laplacians of $\mathcal{G}_1$ and $\mathcal{G}_2$ are:*

$$\mathcal{L}(\mathcal{G}_1) = \begin{pmatrix} 1 & 0 & -1 \\ -1 & 2 & -1 \\ -1 & 0 & 1 \end{pmatrix}, \tag{14}$$

*and*

$$\mathcal{L}(\mathcal{G}_2) = \begin{pmatrix} 1 & 0 & -1 \\ -1 & 1 & 0 \\ -1 & 0 & 1 \end{pmatrix}. \tag{15}$$

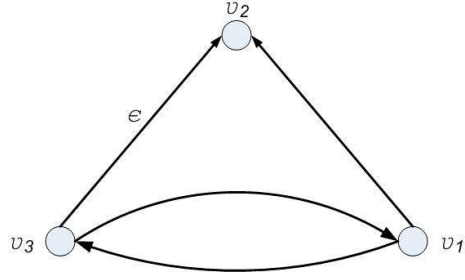

Fig. 1. The above digraph with or without edge $\epsilon$ produce identical dynamics in either of the nodes $\nu_1$ and $\nu_3$.

*The in-degree graph Laplacians in* (14) *and* (15) *can be used to calculate the matrix exponential in* (12) *for* $\mathcal{G}_1$ *and* $\mathcal{G}_2$, *as follows:*

$$e^{-\mathcal{L}(\mathcal{G}_1)t} = \begin{pmatrix} \frac{1}{2}(1+e^{-2t}) & 0 & \frac{1}{2}(1-e^{-2t}) \\ \frac{1}{2}(1-e^{-2t}) & e^{-2t} & \frac{1}{2}(1-e^{-2t}) \\ \frac{1}{2}(1-e^{-2t}) & 0 & \frac{1}{2}(1+e^{-2t}) \end{pmatrix}, \quad (16)$$

*and*

$$e^{-\mathcal{L}(\mathcal{G}_2)t} = \begin{pmatrix} \frac{1}{2}(1+e^{-2t}) & 0 & \frac{1}{2}(1-e^{-2t}) \\ \frac{1}{2}(1-e^{-2t}) & e^{-t} & \frac{1}{2}(1+e^{-2t})-e^{-t} \\ \frac{1}{2}(1-e^{-2t}) & 0 & \frac{1}{2}(1+e^{-2t}) \end{pmatrix}. \quad (17)$$

*It is evident from the expressions in* (16) *and* (17) *that* $\mathcal{G}_1$ *and* $\mathcal{G}_2$ *are distinguishable from* $x_2$, *but not from* $x_1$ *and* $x_3$.

### 3.1. Choice of Initial Conditions

A second look at matrices in (16) and (17) of Example 1 reveals that although the two digraphs are distinguishable from the second agent, any initial conditions $\mathbf{x}(0)$ belonging to the span of the vectors $(1,0,0)^T$ or $(1,1,1)^T$ induces the same response in the second agent. The latter is in fact in the null space of the Laplacian matrix $\mathcal{L}(\mathcal{G})$ for any digraph $\mathcal{G}$; thence,

$$\mathbf{x}(t) = e^{-\mathcal{L}(\mathcal{G})t}\mathbf{x}(0) = \mathbf{0}, \quad (18)$$

for any $\mathbf{x}(0) = \alpha \mathbf{1}, \alpha \in \mathbb{R}$. This gives rise to the interesting question that given two digraphs that are distinguishable from a certain agent what more can be said about the choice of initial conditions leading to non-identical dynamics in the particular agent. The following theorem indicates that the ability to distinguish two digraphs from an agent is a generic property that can be studied independently of the choice of initial condition $\mathbf{x}(0)$.

**Theorem 2** *Given a multi-agent system $\mathcal{S}$ and two distinct digraphs $\mathcal{G}_1 = (\mathcal{V}, \mathcal{E}_1)$ and $\mathcal{G}_2 = (\mathcal{V}, \mathcal{E}_2)$ associated with it, consider a vertex $\nu_i \in \mathcal{V}$ corresponding to agent $x_i \in \mathcal{S}$. If $\mathcal{G}_1$ and $\mathcal{G}_2$ are distinguishable from $x_i$, then for almost all $\mathbf{x}(0) \in \mathbb{R}^{|\mathcal{V}|}$, $x_i^{\mathcal{G}_1}(t) - x_i^{\mathcal{G}_2}(t) \not\equiv 0, t > 0$, where $x_i^{\mathcal{G}_1}(t)$ and $x_i^{\mathcal{G}_2}(t)$ denote the state of agent $x_i$ calculated in the digraphs $\mathcal{G}_1$ and $\mathcal{G}_2$, respectively, with the same initial condition $\mathbf{x}(0)$.*

**Proof.** From the agent response given by (13), it follows that any initial condition $\mathbf{x}(0)$ for which $x_i^{\mathcal{G}_1}(t) - x_i^{\mathcal{G}_2}(t) \equiv 0$, $t > 0$ should satisfy $\boldsymbol{\sigma}(\nu_i)^T(\mathcal{L}(\mathcal{G}_1)^m - \mathcal{L}(\mathcal{G}_2)^m)\mathbf{x}(0) = 0$ for all $m \in \mathbb{N}$. This expression constitutes a set of measure zero in $\mathbb{R}^{|\mathcal{V}|}$, because $\mathcal{G}_1$ and $\mathcal{G}_2$ being distinguishable from $x_i$ implies that there exists an $m \in \mathbb{N}$ such that $\boldsymbol{\sigma}(\nu_i)^T(\mathcal{L}(\mathcal{G}_1)^m - \mathcal{L}(\mathcal{G}_2)^m) \neq 0$. ∎

As a consequence of Theorem 2, any randomly chosen initial conditions will induce different dynamics in an agent from which the digraphs are distinguishable. Of particular interest are the cases, where certain graph symmetries compromise the designer's ability to distinguish between some network failures [18]. One such case is considered in the next example.

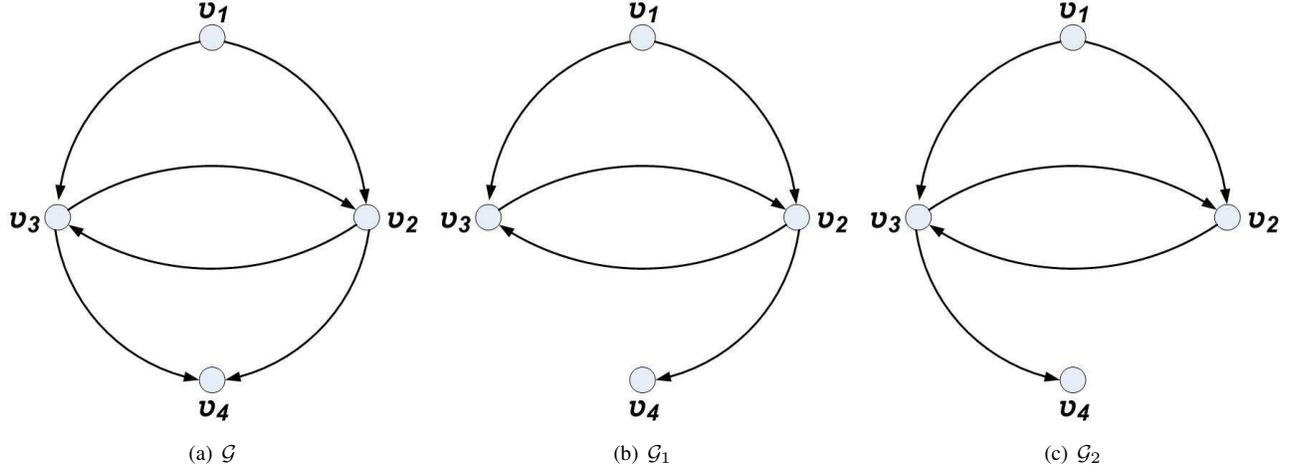

Fig. 2. The digraphs of Example 2.

**Example 2** *Consider a multi-agent system $\mathcal{S} = \{x_1, x_2, x_3, x_4\}$ with the digraph $\mathcal{G} = (\mathcal{V}, \mathcal{E})$, where $\mathcal{V} = \{\nu_1, \nu_2, \nu_3, \nu_4\}$ and $\mathcal{E} = \{(\nu_1, \nu_2), (\nu_1, \nu_3), (\nu_2, \nu_3), (\nu_2, \nu_4), (\nu_3, \nu_2), (\nu_3, \nu_4)\}$, as depicted in Fig. 2(a).*

*The in-degree graph Laplacian for $\mathcal{G}$ is given by:*

$$\mathcal{L}(\mathcal{G}) = \begin{pmatrix} 0 & 0 & 0 & 0 \\ -1 & 2 & -1 & 0 \\ -1 & -1 & 2 & 0 \\ 0 & -1 & -1 & 2 \end{pmatrix}. \tag{19}$$

*Let $\epsilon_1 := (\nu_3, \nu_4)$ and $\epsilon_2 := (\nu_2, \nu_4)$. The digraph of the system after the failure of the communication link between $x_2$ and $x_4$ is $\mathcal{G}_1 = (\mathcal{V}, \mathcal{E}\setminus\{\epsilon_1\})$, and after the failure of the communication link between $x_3$ and $x_4$ is $\mathcal{G}_2 = (\mathcal{V}, \mathcal{E}\setminus\{\epsilon_2\})$, as depicted in Figs. 2(b) and 2(c), respectively.*

*The in-degree graph Laplacians of $\mathcal{G}_1$ and $\mathcal{G}_2$ are:*

$$\mathcal{L}(\mathcal{G}_1) = \begin{pmatrix} 0 & 0 & 0 & 0 \\ -1 & 2 & -1 & 0 \\ -1 & -1 & 2 & 0 \\ 0 & -1 & 0 & 1 \end{pmatrix}, \tag{20}$$

*and*

$$\mathcal{L}(\mathcal{G}_2) = \begin{pmatrix} 0 & 0 & 0 & 0 \\ -1 & 2 & -1 & 0 \\ -1 & -1 & 2 & 0 \\ 0 & 0 & -1 & 1 \end{pmatrix}. \tag{21}$$

*Consider now a permutation $\psi : \mathcal{V} \to \mathcal{V}$ defined as:*

$$\psi(\nu_1) = \nu_1, \ \psi(\nu_2) = \nu_3, \ \psi(\nu_3) = \nu_2, \ \psi(\nu_4) = \nu_4. \tag{22}$$

*The permutation matrix $\Psi$ associated with $\psi(\cdot)$ in (22) is expressed as:*

$$\Psi = \begin{pmatrix} 1 & 0 & 0 & 0 \\ 0 & 0 & 1 & 0 \\ 0 & 1 & 0 & 0 \\ 0 & 0 & 0 & 1 \end{pmatrix}, \tag{23}$$

and since $\Psi \mathcal{L}(\mathcal{G}) = \mathcal{L}(\mathcal{G})\Psi$, hence $\psi(\cdot)$ is an automorphism of $\mathcal{G}$. Moreover, since $\Psi(\Gamma(\nu_3, \nu_4) - \Gamma(\nu_4, \nu_4)) = (\Gamma(\nu_2, \nu_4) - \Gamma(\nu_4, \nu_4))\Psi$, it follows that $\Psi \mathcal{L}(\mathcal{G}_1) = \mathcal{L}(\mathcal{G}_2)\Psi$. In other words, $\epsilon_1$ and $\epsilon_2$ are symmetric under the automorphism given by the permutation matrix $\Psi$. On the other hand, $\Psi\boldsymbol{\sigma}(\nu_1) = \boldsymbol{\sigma}(\nu_1)$ and $\Psi\boldsymbol{\sigma}(\nu_4) = \boldsymbol{\sigma}(\nu_4)$. Therefore, if $\Psi \mathbf{x}(0) = \mathbf{x}(0)$ (i.e. $x_2(0) = x_3(0)$), then both digraphs $\mathcal{G}_1$ and $\mathcal{G}_2$ induce the same responses in either of the agents $x_1$ and $x_4$; that is, $x_1^{\epsilon_1}(t) = x_1^{\epsilon_2}(t)$, $t \geqslant 0$ and $x_4^{\epsilon_1}(t) = x_4^{\epsilon_2}(t)$, $t \geqslant 0$, where the superscripts $\epsilon_i$, $i = 1, 2$ indicate the state values computed in the digraphs $\mathcal{G}_i$, $i = 1, 2$. The reason is that $\mathcal{L}(\mathcal{G}_1) = \Psi^T \mathcal{L}(\mathcal{G}_2)\Psi$, and $\boldsymbol{\sigma}(\nu)^T e^{-\mathcal{L}(\mathcal{G}_1)t} \mathbf{x}(0) = \boldsymbol{\sigma}(\nu)^T e^{-\Psi^T \mathcal{L}(\mathcal{G}_2)\Psi t}\mathbf{x}(0) = \boldsymbol{\sigma}(\nu)^T \Psi^T e^{-\mathcal{L}(\mathcal{G}_2)t}\Psi \mathbf{x}(0)$. Now, since $\boldsymbol{\sigma}(\nu)^T \Psi^T = \boldsymbol{\sigma}(\nu)^T$ for $\nu \in \{2, 4\}$, and $\Psi \mathbf{x}(0) = \mathbf{x}(0)$ because of $x_2(0) = x_3(0)$), agents $x_2$ and $x_4$ will have the same responses in either of the digraphs. It is also worth highlighting that regardless of how the states of the agents are initialized, the temporal evolution of the state of $x_1$ is given by $x_1(t) = x_1(0), t > 0$ and it is not affected by any link failure across the network.

In the following subsection the main theorem of the paper is stated and proved. This theorem provides necessary conditions under which two distinct digraphs, corresponding to a multi-agent system $\mathcal{S}$, are not distinguishable from an agent $x_i \in \mathcal{S}$. It is therefore true that the negation of the stated conditions can be interpreted as sufficient for the digraphs to be distinguishable from $x_i$.

### 3.2. The Main Theorem

The entries of the inverse of the matrix $H$ introduced in Lemma 1 are polynomial fractions in the Laplace variables $s$. Equating the leading coefficients in the $i$−th rows of $H^{-1}$ of two digraphs leads to a necessary consequence for them having identical dynamics in their $i$−th nodes. This is set forth in the following theorem, which constitutes the main result of this section.

**Theorem 3** *Given a multi-agent system $\mathcal{S}$ and two distinct digraphs $\mathcal{G}_1 = (\mathcal{V}, \mathcal{E}_1)$ and $\mathcal{G}_2 = (\mathcal{V}, \mathcal{E}_2)$ associated with it, consider a vertex $\nu_i \in \mathcal{V}$ corresponding to agent $x_i \in \mathcal{S}$. If $\mathcal{G}_1$ and $\mathcal{G}_2$ are not distinguishable from $x_i$, then $\mathrm{d}_1(\nu, \nu_i) = \mathrm{d}_2(\nu, \nu_i)$ for all $\nu \in \mathcal{V}\setminus\{\nu_i\}$, and $\mathrm{c}_k^1(\nu, \nu_i) = \mathrm{c}_k^2(\nu, \nu_i)$ for $k = \mathrm{d}_1(\nu, \nu_i) = \mathrm{d}_2(\nu, \nu_i)$, where $\mathrm{c}_k^l(\nu, \nu_i)$ and $\mathrm{d}_l(\nu, \nu_i)$ denote $\mathrm{c}_k(\nu, \nu_i)$ and $\mathrm{d}(\nu, \nu_i)$, respectively, for digraph $\mathcal{G}_l$, $l = 1, 2$.*

**Proof:** If $\mathcal{G}_1$ and $\mathcal{G}_2$ are not distinguishable from $x_i$, then $x^{\mathcal{G}_1}(t) - x^{\mathcal{G}_2}(t) \equiv 0$ for any $\mathbf{x}(0) \in \mathbb{R}^{|\mathcal{V}|}$, and from (13), it follows that:

$$\boldsymbol{\sigma}(\nu_i)^T e^{-\mathcal{L}(\mathcal{G}_1)t} = \boldsymbol{\sigma}(\nu_i)^T e^{-\mathcal{L}(\mathcal{G}_2)t}, \ t > 0. \tag{24}$$

Let $H_l = sI + \mathcal{L}(\mathcal{G}_l)$, $l = 1, 2$, where $s \in \mathbb{C}$ is the Laplace variable. Taking the Laplace transform of (24):

$$\boldsymbol{\sigma}(\nu_i)^T H_1^{-1} = \boldsymbol{\sigma}(\nu_i)^T H_2^{-1}, \tag{25}$$

or equivalently:

$$\left[H_1^{-1}\right]_{ip} = \left[H_2^{-1}\right]_{ip}, \ p \in \mathbb{N}_{|\mathcal{V}|}. \tag{26}$$

The adjoint of $H_l$, $l = 1, 2$ can now be used to compute the inverse matrices as follows:

$$\left[H_l^{-1}\right]_{nm} = \frac{(-1)^{m+n} \det\left(C_{mn}^l\right)}{\det\left(H_l\right)}, \ m, n \in \mathbb{N}_{|\mathcal{V}|}, \ l = 1, 2, \tag{27}$$

where $C_{mn}^l$, $l = 1, 2$, is the matrix obtained by removing the $m$−th row and $n$−th column of $H_l$, $l = 1, 2$. It follows from (26) and (27) that:

$$\frac{\det\left(C_{ji}^1\right)}{\det\left(H_1\right)} = \frac{\det\left(C_{ji}^2\right)}{\det\left(H_2\right)}, j \in \mathbb{N}_{|\mathcal{V}|}, \tag{28}$$

or:

$$\det\left(C_{ji}^1\right)\det\left(H_2\right) = \det\left(C_{ji}^2\right)\det\left(H_1\right), j \in \mathbb{N}_{|\mathcal{V}|}. \tag{29}$$

Next, $\det(H_l), l = 1, 2$ in (29) can be written as a summation over the product of all possible permutations on $|\mathcal{V}|$ matrix elements [11], as given below:

$$\det(H_2) = \sum_{\psi \in \Xi_{|\mathcal{V}|}} (-1)^{\iota(\psi)} \prod_{j=1}^{|\mathcal{V}|} [H_2]_{j\psi(j)}, \tag{30a}$$

$$\det(H_1) = \sum_{\psi \in \Xi_{|\mathcal{V}|}} (-1)^{\iota(\psi)} \prod_{j=1}^{|\mathcal{V}|} [H_1]_{j\psi(j)}, \tag{30b}$$

where $\Xi_{|\mathcal{V}|}$ is the finite group formed by the $(|\mathcal{V}|)!$ permutations on the set $\mathbb{N}_{|\mathcal{V}|}$. Moreover, for a permutation $\psi(.)$ on the set $\mathbb{N}_{|\mathcal{V}|}$, $\iota(\psi)$ denotes its length. For $l = 1, 2$, note that the only term in $\det(H_l)$ which includes $s^{|\mathbb{V}|}$ (or $s^{|\mathbb{V}|-1}$) in the right-hand sides of (30) is the one corresponding to the diagonal elements of $H_l$, and is given by:

$$\prod_{k=1}^{|\mathcal{V}|} [H_l]_{kk} = \prod_{\nu \in \mathcal{V}} (s + d_{in}^l(\nu)) = s^{|\mathcal{V}|} + \left( \sum_{\nu \in \mathcal{V}} d_{in}^l(\nu) \right) s^{|\mathcal{V}|-1} + \ldots + \prod_{\nu \in \mathcal{V}} d_{in}^l(\nu), \tag{31}$$

where for $\nu \in \mathcal{V}$ and $l = 1, 2$, $d_{in}^l(\nu)$ denotes $d_{in}(\nu)$ in $\mathcal{G}_l$. The power of $s$ in any other term in the right-hand sides of (30) is less than $|\mathbb{V}| - 1$. The proof now follows immediately from Lemma 1 and upon equating the leading coefficients and the polynomial degrees in the left-hand side and right-hand side of (29) for $j \in \mathbb{N}_{|\mathcal{V}|} \backslash \{i\}$. ∎

According to Theorem 3, in order to determine that two digraphs are distinguishable from an agent corresponding to vertex $\nu_i$ it suffices to find some vertex whose distance to $\nu_i$ differs in each digraph, or else if all vertices are at the same distance from $\nu_i$ in either digraphs, then it is also sufficient to find a vertex who has a different number of shortest paths to vertex $\nu_i$ in each digraph. The authors in [1] provide relevant claims that connect the relative degree and gain factor of the transfer function of a consensus network with an observer node (output) and a controller node (input) to the length and number of shortest paths between those two nodes. However, the proof they provide is not correct because in Lemma 3.1 they make use of a matrix product commutative property that does not hold true.

The next corollary follows by applying the sufficient conditions of Theorem 3 to the tail vertices of $\partial_{\mathcal{G}}^- \{\nu\}$ for an observing agent $\{\nu_i\}$, and provides useful information about indistinguishable digraphs. In essence, it states that such digraphs should be identical in the immediate (incoming) vicinity of the observation point.

**Corollary 1** *Consider a multi-agent system $\mathcal{S}$, an agent $x \in \mathcal{S}$, and two distinct digraphs $\mathcal{G}_1 = (\mathcal{V}, \mathcal{E}_1)$ and $\mathcal{G}_2 = (\mathcal{V}, \mathcal{E}_2)$ that are both associated with $\mathcal{S}$. Let $\nu_i \in \mathcal{V}$ be the vertex that corresponds to $x$, for some $i \in \mathbb{N}_{|\mathcal{V}|}$. If $\mathcal{G}_1$ and $\mathcal{G}_2$ are not distinguishable from $x$, then $\boldsymbol{\sigma}(\nu_i)\mathcal{L}(\mathcal{G}_1) = \boldsymbol{\sigma}(\nu_i)\mathcal{L}(\mathcal{G}_2)$, so that $\partial_{\mathcal{G}_1}^-\{\nu_i\} = \partial_{\mathcal{G}_2}^-\{\nu_i\}$, and in particular $d_{\mathcal{G}_1}^-\{\nu_i\} = d_{\mathcal{G}_2}^-\{\nu_i\}$.*

## IV. JOINTLY DETECTABLE LINK FAILURES

In this section, the notion of distinguishable digraphs from Definition 1 is used to define and characterize the concept of *joint detectability* for a multi-agent system subject to simultaneous link failures.

**Definition 2** *Given a multi-agent system $\mathcal{S}$ and its associated digraph $\mathcal{G} = (\mathcal{V}, \mathcal{E})$, a subset of edges $\mathcal{E}_1 \subset \mathcal{E}$ is said to be jointly detectable from the agent $x \in \mathcal{S}$ if $\mathcal{G}$ and $\mathcal{G}_1 = (\mathcal{V}, \mathcal{E} \backslash \mathcal{E}_1)$ are distinguishable from $x$.*

The next Proposition follows from Theorem 3 using a contradiction argument. The proposition provides sufficient graphical conditions for joint detectability of a link-set from an observing agent.

**Proposition 1** *Given a multi-agent system $\mathcal{S}$ and its associated digraph $\mathcal{G}_1 = (\mathcal{V}, \mathcal{E})$, consider a vertex $\nu_i \in \mathcal{V}$ corresponding to agent $x_i \in \mathcal{S}$, and a subset of edges $\mathcal{E}_1 \subset \mathcal{E}$. If there exists an edge $\epsilon := (\nu_j, \nu_k) \in \mathcal{E}_1$ such that $d(\nu_j, \nu_i) > d(\nu_k, \nu_i)$, then $\mathcal{E}_1$ is jointly detectable from $x_i$.*

**Proof:** The proof follows by contradiction. Suppose that $\mathcal{E}_1$ is not jointly detectable from $x_i$ and denote $\mathcal{G}_2 = (\mathcal{V}, \mathcal{E} \backslash \mathcal{E}_1)$, such that $\mathcal{G}_1$ and $\mathcal{G}_2$ are not distinguishable from $x_i$.

It follows from Theorem 3 that for some $\{\hat{d}_1, \hat{d}_2\} \subset \mathbb{N}_{|\mathcal{V}|-2}$:

$$\hat{d}_1 = \mathrm{d}_1(\nu_k, \nu_i) = \mathrm{d}_2(\nu_k, \nu_i), \tag{32a}$$

$$\hat{d}_2 = \mathrm{d}_1(\nu_j, \nu_i) = \mathrm{d}_2(\nu_j, \nu_i), \tag{32b}$$

and

$$\mathrm{c}^1_{\hat{d}_1}(\nu_k, \nu_i) = \mathrm{c}^2_{\hat{d}_1}(\nu_k, \nu_i), \tag{33a}$$

$$\mathrm{c}^1_{\hat{d}_2}(\nu_j, \nu_i) = \mathrm{c}^2_{\hat{d}_2}(\nu_j, \nu_i), \tag{33b}$$

where for $\{\tau, \nu\} \subset \mathcal{V}$ and $l = 1, 2$, $\mathrm{c}^l_k(\tau, \nu)$ and $\mathrm{d}_l(\tau, \nu)$ denote $\mathrm{c}_k(\tau, \nu)$ and $\mathrm{d}(\tau, \nu)$, respectively, for digraph $\mathcal{G}_l$.

Let $\mathcal{P}$ be a $\nu_k \nu_i$ path with length $\hat{d}_1 = \mathrm{d}_2(\nu_k, \nu_i)$ in $\mathcal{G}_2$. Since $\epsilon \mathcal{P}$ is a $\nu_j \nu_i$ path with length $1 + \hat{d}_1$ in $\mathcal{G}_1$, one has that:

$$\mathrm{d}_1(\nu_j, \nu_i) \leqslant 1 + \mathrm{d}_2(\nu_k, \nu_i). \tag{34}$$

On the other hand, the assumption $\mathrm{d}_1(\nu_k, \nu_i) < \mathrm{d}_1(\nu_j, \nu_i)$, together with (32a), leads to:

$$\mathrm{d}_2(\nu_k, \nu_i) < \mathrm{d}_1(\nu_j, \nu_i). \tag{35}$$

The inequalities (34) and (35) imply that:

$$\mathrm{d}_1(\nu_j, \nu_i) = 1 + \mathrm{d}_2(\nu_k, \nu_i). \tag{36}$$

Next, from (32) and (36) it follows that $\hat{d}_2 = 1 + \hat{d}_1$ and note that since $(\mathcal{E} \backslash \mathcal{E}_1) \subset \mathcal{E}$, every $\nu_j \nu_i$ path with length $\hat{d}_2$ in $\mathcal{G}_2$ is a $\nu_j \nu_i$ path with length $\hat{d}_2$ in $\mathcal{G}_1$, while $\epsilon \mathcal{P}$ is a $\nu_j \nu_i$ path with length $\hat{d}_1 + 1 = \hat{d}_2 = \mathrm{d}_1(\nu_j, \nu_i)$ in $\mathcal{G}_1$ that does not exist in $\mathcal{G}_2$. Hence, $\mathrm{c}^1_{\hat{d}_2}(\nu_j, \nu_i) \geqslant 1 + \mathrm{c}^2_{\hat{d}_2}(\nu_j, \nu_i)$, which is in contradiction with (33b). ∎

In particular, $\mathrm{d}(\nu_k, \nu_i)$ in Proposition 1 can be equal to zero, in which case $\epsilon \in \partial^-_{\mathcal{G}_1}\{\nu_i\}$ and any subset containing such an edge will be jointly detectable from $x_i$. Intuitively, under the conditions specified in Proposition 1 the vertex $\nu_j$ is located at a greater distance from the observing vertex $\nu_i$ as compared to the vertex $\nu_k$. Hence, the *"information"* concerning the state of $\nu_j$ will *"reach"* the observing agent *"faster"* when passed through the failed edge $(\nu_j, \nu_k)$. The next proposition extends the result provided in [17] for a single link to any combination of links with a common head vertex.

**Proposition 2** *Consider a multi-agent system $\mathcal{S}$, its associated digraph $\mathcal{G}_1 = (\mathcal{V}, \mathcal{E})$, an agent $x_i \in \mathcal{S}$ with its corresponding vertex $\nu_i \in \mathcal{V}$, a vertex $\nu_k \in \mathcal{V}$, and a subset of edges $\mathcal{E}_2 \subset \partial^-_{\mathcal{G}_1}\{\nu_k\}$. If there exists an out-branching root $\nu_o \in \mathcal{V} \backslash \{\nu_k\}$ as well as a $\nu_k \nu_o$ path in $\mathcal{G}_2 = (\mathcal{V}, \mathcal{E} \backslash \mathcal{E}_2)$, then $\mathcal{E}_2$ is jointly detectable from $x_i$.*

**Proof:** The result for $k = i$ is a direct consequence of (10). For $k \neq i$ suppose that Proposition 2 does not hold true, i.e., $\nu_o$ is an out-branching root of $\mathcal{G}_2$ and there exists a $\nu_k \nu_o$ path in $\mathcal{G}_2$, but $x^{\mathcal{G}_1}(t) - x^{\mathcal{G}_2}(t) \equiv 0$ for all $\mathbf{x}(0) \in \mathbb{R}^{|\mathcal{V}|}$, where $x^{\mathcal{G}_1}(t)$ and $x^{\mathcal{G}_2}(t)$ denote the state of the agent $x$ calculated in the digraphs $\mathcal{G}_1$ and $\mathcal{G}_2$, respectively, with the same set of initial conditions $\mathbf{x}(0)$. Let $H_l = sI + \mathcal{L}(\mathcal{G}_l)$, $l = 1, 2$, where $s \in \mathbb{C}$ is the Laplace variable. Moreover, denote the $(|\mathcal{V}| - 1) \times (|\mathcal{V}| - 1)$ sub-matrix that results from removing the $m$–th row and $n$–th column of $H_l$ by $C^l_{mn}$, $l = 1, 2$. Since $x^{\mathcal{G}_1}(t) - x^{\mathcal{G}_2}(t) \equiv 0$ for all $\mathbf{x}(0) \in \mathbb{R}^{|\mathcal{V}|}$, (24) to (27) still hold and it follows from the equality:

$$\mathcal{L}(\mathcal{G}_2) = \mathcal{L}(\mathcal{G}_1) - |\mathcal{E}_2| \Gamma((\nu_k, \nu_k)) + \sum_{\epsilon \in \mathcal{E}_2} \Gamma(\epsilon), \tag{37}$$

that the matrices $H_l$, $l = 1, 2$, will only differ at $[H_l]_{kk}$ and $[H_l]_{kj}$ for all $j \in \{n \in \mathbb{N}_{|\mathcal{V}|}; (\nu_n, \nu_k) \in \mathcal{E}_2\}$. Therefore, after removing the $k$–th row from $H_l$, $l = 1, 2$, one has that $C^1_{ki} = C^2_{ki}$, $s \in \mathbb{C}$, or $\det(C^1_{ki}) \equiv \det(C^2_{ki}) \not\equiv 0$, where the inequality follows from Theorem 1, since $\det(C^2_{ki})$ at $s = 0$ is equal to the number of $\nu_k$-rooted spanning out-branchings of $\mathcal{G}_2$. On the other hand, it follows from (26) that $\left[H_1^{-1}\right]_{ik} \equiv \left[H_2^{-1}\right]_{ik}$. Now, substituting from (27) and using $\det(C^1_{ki}) \equiv \det(C^2_{ki}) \not\equiv 0$, one arrives at the relation $\det(sI + \mathcal{L}(\mathcal{G}_1)) \equiv \det(sI + \mathcal{L}(\mathcal{G}_2))$. Similarly, the equality $[H_1^{-1}]_{io} \equiv [H_2^{-1}]_{io}$ along with the preceding result yields $\det(C^1_{oi}) = \det(C^2_{oi})$, $s \in \mathbb{C}$. According to Theorem 1, for $s = 0$ this means that $\mathcal{G}_1$ and $\mathcal{G}_2$ have the same number of $\nu_o$-rooted spanning out-branchings. This, however, is a contradiction since $\nu_o$ being an out-branching root of $\mathcal{G}_2$, together with $\nu_k \neq \nu_o$, implies that the number of $\nu_o$-rooted spanning out-branchings in $\mathcal{G}_1$ is strictly greater than $\mathcal{G}_2$. ∎

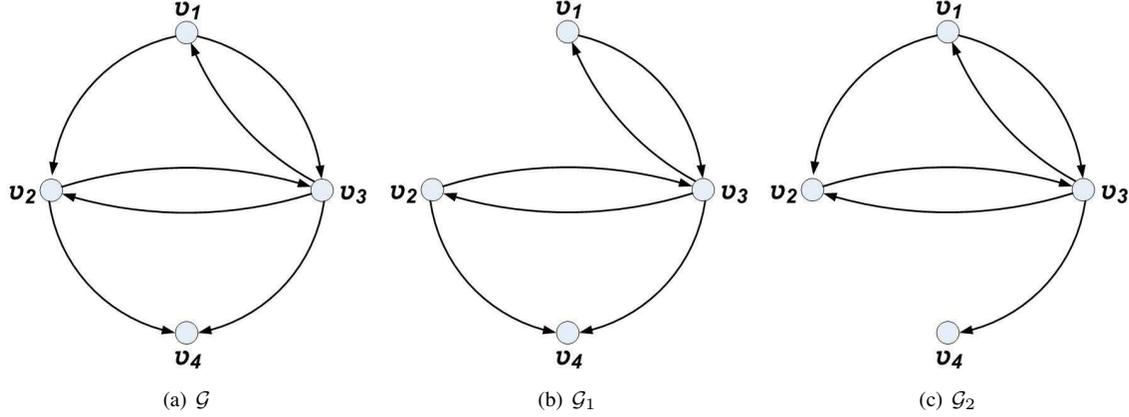

Fig. 3. The digraphs for Example 3.

It is notable that if all of the links in the set $\partial^-_{\mathcal{G}_1}\{\nu_k\}$ in Proposition 2 are simultaneously removed to form the digraph $\mathcal{G}_2$, then the vertex $\nu_o \neq \nu_k$ is not an out-branching root of $\mathcal{G}_2$. Therefore, the set $\partial^-_{\mathcal{G}_1}\{\nu_k\}$ does not satisfy the sufficient conditions given by Proposition 2 for joint detectability from agent $x_i$. It is worth highlighting that under the conditions of Proposition 2, $\nu_k$ is also an out-branching root of $\mathcal{G}$, so that there exists a $\nu_k\nu_i$ path from the head vertex of the failed links to the vertex corresponding to the observing agent $x_i$. Consequently, the *"information"* concerning the link failures can *"reach"* the observing agent. Even though the sufficient conditions in Proposition 2 may be too conservative, they do pave the way for the following useful Corollary for the special case of strongly connected digraphs, of which connected undirected graphs can be regarded as a special class.

**Corollary 2** *Consider a multi-agent system $\mathcal{S}$ and its associated digraph $\mathcal{G}_1 = (\mathcal{V}, \mathcal{E})$. If $\mathcal{G}_2 = (\mathcal{V}, \mathcal{E}\setminus\{\epsilon\})$ is strongly connected for all $\epsilon \in \mathcal{E}$, then any edge $\{\epsilon\}$ is jointly detectable from any agent $x$, $\forall \epsilon \in \mathcal{E}$ and $\forall x \in \mathcal{S}$.*

**Example 3** *Consider a multi-agent system $\mathcal{S} = \{x_1, x_2, x_3, x_4\}$ with the digraph $\mathcal{G} = (\mathcal{V}, \mathcal{E})$, where $\mathcal{V} = \{\nu_1, \nu_2, \nu_3, \nu_4\}$ and $\mathcal{E} = \{(\nu_1, \nu_2), (\nu_1, \nu_3), (\nu_2, \nu_3), (\nu_2, \nu_4), (\nu_3, \nu_1), (\nu_3, \nu_2), (\nu_3, \nu_4)\}$, as depicted in Fig. 3(a). The digraphs resulting from the failure of each of the links $\epsilon_1 := (\nu_1, \nu_2)$ and $\epsilon_4 := (\nu_2, \nu_4)$ are $\mathcal{G}_1 = (\mathcal{V}, \mathcal{E}\setminus\{\epsilon_1\})$ and $\mathcal{G}_4 = (\mathcal{V}, \mathcal{E}\setminus\{\epsilon_4\})$, which are depicted in Figs. 3(b) and 3(c). The responses of $x_1$ calculated in digraphs $\mathcal{G}$, $\mathcal{G}_1$ and $\mathcal{G}_2$ are denoted by $x_1(t)$, $x_1^{\epsilon_1}(t)$, and $x_1^{\epsilon_4}(t)$, respectively, and they are depicted in Figs. 4(a) and 4(b), with the same random initial states. It is straightforward to verify that $\epsilon_4$ is not detectable from $x_1$ because although $\nu_1$ is an out-branching root, $\nu_4$ is not and there does not exist a path from $\nu_4$ to $\nu_1$ in $\mathcal{G}_2$.*

**Remark 1** *For the cases where the agents are also prone to failures, it is notable that the effect of the removal of a node from the digraph $\mathcal{G}$ is equivalent to removal of all edges that are incident to that node. It is therefore true that the failure of the agent corresponding to node $\nu$ is detectable from agent $x$ if the set of edges $\partial^+_{\mathcal{G}}\{\nu\} \cup \partial^-_{\mathcal{G}}\{\nu\}$ is jointly detectable from agent $x$.*

To close this section, it is worth noting that the proof for the sufficient conditions in Proposition 1 relies on the highest power of $s$ and the leading coefficients in the minors of the transformed graph Laplacian. According to Lemma 1, these parameters depend on the length and number of shortest paths between the graph vertices. On the other hand, the proof of Proposition 2 revolves around the minors of the transformed graph Laplacian at $s = 0$, which according to Theorem 1 are related to the number of rooted spanning out-branchings in the digraph. The two sets of results can therefore be interpreted as the two extremes corresponding to $s \to \infty$ (leading coefficients) and $s = 0$ (constant terms) in the minors of the transformed graph Laplacian (which are polynomials in $s$). The next subsection offers a complete graphical representation for the class of jointly detectable links that satisfy the condition of Proposition 1.

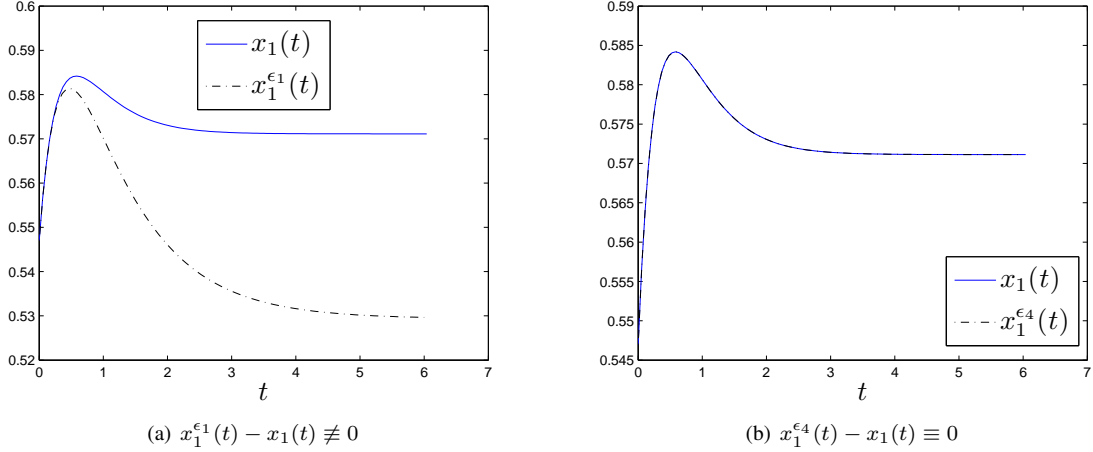

Fig. 4. The responses of $x_1$ in digraphs of Example 3.

### 4.1. Shortest Paths Subgraph

The next definition and the corollary which follow provide a complete characterization of the class of detectable links that satisfy the sufficient condition given in Proposition 1.

**Definition 3** *Given a digraph $\mathcal{G} = (\mathcal{V}, \mathcal{E})$ and a vertex $\nu \in \mathcal{V}$, suppose that $\forall \mu \in \mathcal{V}\setminus\{\nu\}$, there exists a $\mu\nu$ path in $\mathcal{G}$. The shortest subgraph of $\mathcal{G}$ w.r.t. $\nu$ is identified by $\mathcal{K}_\mathcal{G}^\nu = (\mathcal{V}, \mathcal{E}_{\mathcal{K},\nu})$, where $\mathcal{E}_{\mathcal{K},\nu} \subset \mathcal{E}\setminus\partial_\mathcal{G}^+\{\nu\}$ and $\forall \epsilon := (\tau, \vartheta) \in \mathcal{E}\setminus\partial_\mathcal{G}^+\{\nu\}$, $\epsilon \in \mathcal{E}_{\mathcal{K},\nu} \longleftrightarrow \mathrm{d}(\tau,\nu) = \mathrm{d}(\vartheta,\nu) + 1$.*

Any link $(\tau, \vartheta) \in \mathcal{E}\setminus\partial_\mathcal{G}^+\{\nu\}$ that satisfies the condition $d(\tau,\nu) = d(\vartheta,\nu) + 1$ in Definition 3 is guaranteed to contribute to a shortest path from $\tau$ to $\nu$. In particular, such a link satisfies the sufficient condition for detectability from $\nu$ as expressed in Proposition 1. The following corollary is a restatement of Proposition 1 in terms of Definition 3.

**Corollary 3** *Given a multi-agent system $\mathcal{S}$ and its associated digraph $\mathcal{G} = (\mathcal{V}, \mathcal{E})$, consider a vertex $\nu \in \mathcal{V}$ corresponding to agent $x \in \mathcal{S}$, and a subset of edges $\mathcal{E}_f \subset \mathcal{E}$. Assume also that for any $\mu \in \mathcal{V}\setminus\{\nu\}$, there exists a $\mu\nu$ path in $\mathcal{G}$, and let $\mathcal{K}_\mathcal{G}^\nu = (\mathcal{V}, \mathcal{E}_{\mathcal{K},\nu})$ denote the shortest paths subgraph of $\mathcal{G}$ w.r.t. $\nu$. If $\mathcal{E}_f \cap \mathcal{E}_{\mathcal{K},\nu} \neq \varnothing$, then $\mathcal{E}_f$ is jointly detectable from $x$.*

### 4.2. Multiple Observation Points

If the states of more than one agent are available to the designer, then two graphs will be distinguishable from a subset of nodes if and only if they are distinguishable in the sense of Definition 1, from at least one of the accessible agents. Similarly, a given subset of links is detectable from a subset of agents if and only if the link-set is detectable from at least one of the available agents. The concept of shortest path subgraph from the previous subsection can be used to derive the following greedy heuristic, which is a computationally efficient procedure for determining a set $\mathcal{O}$ of nodes from which the failure of any subset of links in the digraph will be detectable.

**Procedure 1** Determine Set $\mathcal{O}$ of Observation Points

**Input:** $\mathcal{G} = (\mathcal{V}, \mathcal{E})$
1: $\mathcal{O} \Leftarrow \varnothing$
2: **while** $\mathcal{E} \neq \varnothing$ **do**
3: $\quad \nu \Leftarrow \arg\max\{|\mathcal{E}_{\mathcal{K},\hat{\nu}}|; \hat{\nu} \in \mathcal{V}\}$
4: $\quad \mathcal{O} \Leftarrow \mathcal{O} \cup \nu$
5: $\quad \mathcal{E} \Leftarrow \mathcal{E} \backslash \mathcal{E}_{\mathcal{K},\nu}$
6: **end while**
**Output:** $\mathcal{O}$

## V. CONCLUSIONS

The concept of distinguishable digraphs provides an effective mathematical characterization for the simultaneous failure of links in multi-agent systems. The results indicate that the ability to distinguish two digraphs from a given agent is a generic property that can be studied independently of the choice of initial states. The analytical developments in this paper include a powerful extension to the all-minors matrix tree theorem that relates the polynomial degree and leading coefficient of the minors of the transformed Laplacian of a digraph to the length and number of shortest paths connecting its vertices. Moreover, the analytical results reveal an intricate relationship between the ability to distinguish between the dynamic response of two multi-agent systems and the inter-nodal distances and number of shortest paths in their digraphs. Detectability of link failures is studied as a special case, and sufficient conditions indicate that, in order for a group of links to be jointly detectable from a given agent, it suffices to design the information flow digraph in such a way that for one of the links in the group, the head vertex is at a shorter topological distance from the observing agent, as compared to the tail vertex. Alternatively, a group of links with a common head are jointly detectable from any agent in the network, provided the common head vertex is an out-branching root and there exists another out-branching root in the network information flow digraph. The concept of shortest paths subgraph can be used to efficiently select a subset of nodes from which the failure of any set of links is detectable. Developing efficient detection and isolation procedures is the next step for future research.